# Mesoscopic scale study of lateral dynamics of Sn-intercalation of the buffer layer on SiC

Benno Harling[1], Zamin Mamiyev[2], Christoph Tegenkamp[2], Martin Wenderoth[1]

[1] IV. Physikalisches Institut, Georg-August-Universität, Friedrich-Hund-Platz 1, 37077 Göttingen, Germany

[2] Institut für Physik, Technische Universität Chemnitz, Reichenhainer Str. 70, 09126 Chemnitz, Germany

E-mail: benno.harling@uni-goettingen.de

martin.wenderoth@uni-goettingen.de

Abstract

The dynamics of Sn intercalation of the buffer layer on SiC was investigated with a frozen-in diffusion front using Kelvin Probe Force Microscopy. The technique allows to laterally distinguish between intercalated regions and the pristine buffer layer. Comparing topography features with the surface potential on the mesoscopic scale confirms that surface steps act as transport barriers. The results show a distinct pin hole mechanism for the material flow from terrace to terrace of the vicinal substrate surface. Nucleation and the formation of tin intercalated phase on terraces happen at steps. This results in a mesoscopic growth against the macroscopic diffusion direction.

Introduction

Intercalation of two-dimensional materials, i.e., the process of inserting atoms or molecules in between the layers, is a rather old topic which has regained recent interest due to its high relevance in modern technologies like, e.g., battery charging [1]. Moreover, intercalation opens a path toward steering the electronic properties of the layered system itself by utilizing the proximity of two very different systems, similar to the idea of stacking layers of different van-der-Waals materials [2]. Epitaxial graphene (EG) grown on silicon carbide (SiC) provides a suitable platform to study intercalation on different length scales.

Starting from the carbon buffer layer on SiC, this process can be used for a variety of atoms. EG on SiC(0001) has an intrinsic Fermi energy ($E_F$) $\approx 400$ meV above the Dirac point of graphene ($E_D$). The filling of the Dirac cone, i.e., the doping level, can be changed by intercalation, e.g., by hydrogen, it shifts the $E_F$ to $\approx -100$ meV resulting in an effective p-type doping. [3]. Intercalation was already performed with a vast number of elements besides hydrogen, e.g. Sn [4], Ga [5], Yb [6], Pb [7, 8], Tb [9], S [10], Sb [11], Gd [12], Ag [13], Ru [14] and others.

To finally have a quasi-free monolayer graphene (QFMLG), most intercalation recipes start from the BL on top of the SiC crystal. While carbon in graphene is entirely $sp^2$ hybridized, the BL represents a carbon system with a graphene structure that consists of $sp^2$- as well as $sp^3$-bonds, reflecting the partly covalent bonding to the SiC substrate [15]. During the intercalation process the BL is transferred to graphene, ending up in a van-der-Waals bonded QFMLG, while the SiC-bonds are saturated by the intercalated atom species.

While many of these studies investigate the impact of intercalation on the electronic properties of the host material EG, more recently, a series of papers moved the focus to the atomic-scale processes of intercalation in graphene itself for SiC and other substrate systems, including both experimental and theoretical studies. They have addressed (1) the transport of atoms from the surface site to an intercalated one, (2) the process of transport of atoms, as well as (3) local diffusion barriers [5].[16]

While these results provide a first picture, the challenging question, how the lateral transport, the bonding of the foreign atom and the decoupling of the BL from the substrate is interacting to eventually form an epitaxial graphene layer, still remains open.

To further investigate the impact of the substrate steps on the material flow, we study the diffusion process by heating the sample to the intercalation regime and cooling it down to ambient temperature. With this, the diffusion process is frozen-in. Kelvin Probe Force Microscopy (KPFM) gives us sensitivity to discriminate between the pristine and intercalated areas of the surface. Eventually, the simultaneously acquired topographies allow us to study the 2D-diffusion process on a mesoscopic scale. Substrate steps turn out to be dominant for the material flow of the tin, including nucleation and 2D growth.

Methods

Sample Preparation

The intercalated graphene buffer layer samples used within this study were grown on the Si terminated SiC(0001)-surface of 4H n-type doped silicon carbide. The PASG method [17] was used to get a homogeneous buffer layer with a low amount of graphene.

A few monolayers of tin were deposited onto the BL sample at room temperature using a shadow mask technique. For intercalation, the sample was heated up to 800 °C, resulting in a Sn(1×1) phase with respect to the substrate [18]. Afterward, the temperature was increased to 1050 °C to form ($\sqrt{3} \times \sqrt{3}$) interface reconstruction. This last step comes along with a partial de-intercalation of the Sn, which is required for ($\sqrt{3} \times \sqrt{3}$) formation at 1/3 ML local coverage. Since the perfect Sn(1×1) interface is highly robust (or it is a stable phase) against temperature, the de-intercalation of 2/3 ML of Sn is highly determined by the imperfections of the interface, in addition to the annealing temperature and time [19]. Therefore, in most cases, the remaining patches limit lateral extent of the ($\sqrt{3} \times \sqrt{3}$) phase, resulting in mixed ($\sqrt{3} \times \sqrt{3}$) and Sn(1×1) phases [20], as we will also discuss in the following.

KPFM Setup

Kelvin Probe Force Microscopy (KPFM) measurements utilize the combination of a conventional AM-AFM (Agilent 5600SL) in intermitted contact mode at ambient conditions with the commercially available upgrade to observe the surface's electric potential with the KPFM-mode as described by [21].

A conducting Al/Pt tip and a second feedback loop with a lock-in amplifier is used, applying a bias voltage to the tip, canceling out the difference between the local electric potential of the sample and the tip. The applied bias voltage is the negative electric potential difference also called contact potential difference (CPD) of the tip and sample.

Without calibration, the absolute value cannot be estimated, as the tip's electric potential is dependent on its unknown apex configuration. Relative values within one measurement are still valid as long as the tip does not change its configuration during scanning. This can easily be recognized, as tip modifications result mostly in big changes of absolute value, which is easily noticeable and should not be a concern within these measurements.

Results

The Sn intercalation process and related phase transformations were investigated in situ using high-resolution Spot Profile Analysis-Low Energy Electron Diffraction (SPA-LEED). SPA-LEED, with its exceptionally high resolution and enhanced transfer width of approximately 200 nm, enables precise analysis of nanoscale periodicities and surface roughness, as well as direct access to diffraction profiles. [22]

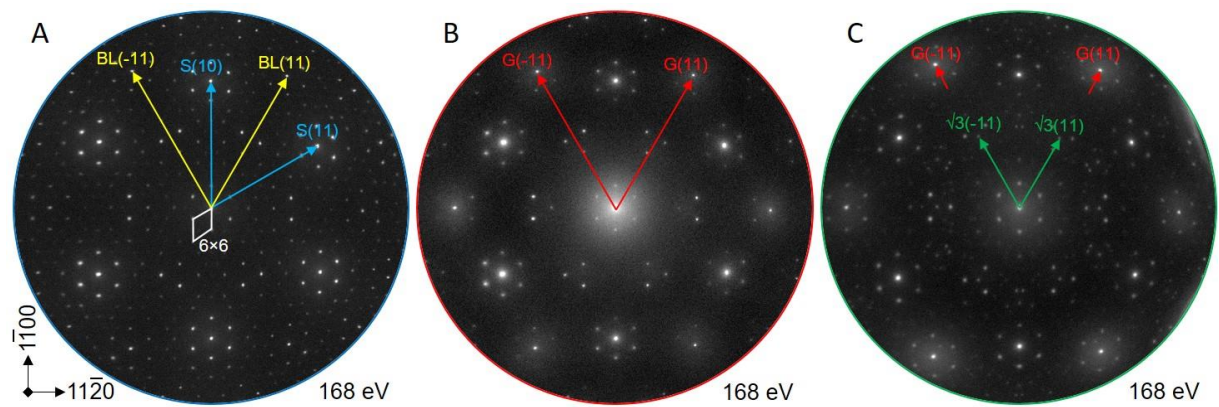

Figure 1. SPA-LEED images for the clean buffer layer (A), after Sn(1×1) intercalation (B) and following partial phase transformation to a ($\sqrt{3} \times \sqrt{3}$)R30° (C). The first-order SiC (S), a buffer layer (BL), and graphene (G) vectors are marked.

In Fig. 1A, we present a high-resolution SPA-LEED image of the clean buffer layer (BL), which shows clear ($6\sqrt{3} \times 6\sqrt{3}$)R30° (referred to as $6\sqrt{3}$ in the following) with respect to the SiC(0001) substrate. As shown in Fig. 1B, the diffraction pattern changes after Sn intercalation, with the new surface displaying brighter Gr(1×1) spots accompanied by a broad, coherent background and suppressed $6\sqrt{3}$ periodicity. These changes confirm the success of the Sn intercalation process and the decoupling of graphene from the substrate. [18] We identify this phase as EG/Sn(1×1), indicating that the SiC(1×1) surface is fully saturated with intercalated Sn atoms, forming a local monolayer coverage. Upon annealing this phase at 1050°C, the intensity of the Gr(1×1) spots and the coherent background decrease, while the $6\sqrt{3}$ periodicity reappears, as seen in Fig. 1C. This change is attributed to partial deintercalation of the Sn layer, resulting in recovery of the BL. Notably, the resulting surface displays a ($\sqrt{3} \times \sqrt{3}$)R30° periodicity. With respect to the SiC substrate. [20]

We identify reconstructions at the tin deposited areas of the sample to be mainly (1×1) and a fraction of $(\sqrt{3} \times \sqrt{3})R30°$ surface reconstruction with respect to the substrate. The ratio of both varies as a function of position, most probably due to a temperature gradient during preparation. Moving from position α to the shadow mask (indicated by the dashed line) and β in Fig. 2 A, the fraction of (1x1) relative to $(\sqrt{3} \times \sqrt{3})R30°$ increase.

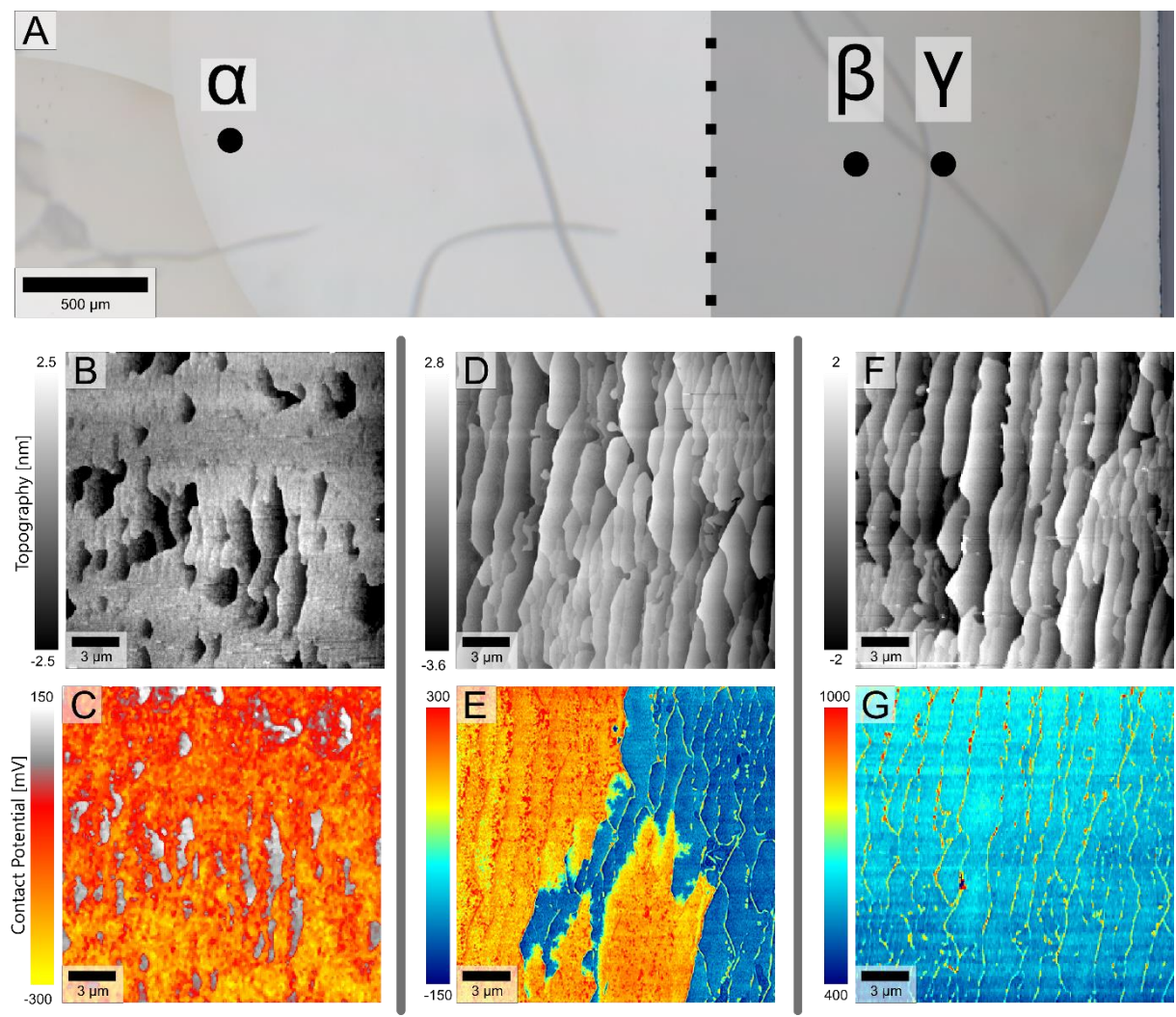

Figure 2. A shows an optical image of the sample. The dotted line indicates the edge of the mask. While the left side was deposited with few Sn monolayers, the right side (gray shaded area) was intentionally unexposed. α, β, and γ indicate the position where corresponding measurements were taken. B/C, D/E and F/G are combined data sets showing the surface topography (AFM) that were taken simultaneously with the contact potential difference maps (CPD).

In the following, we will discuss the transition from the (1×1) reconstruction (position α) to the undeposited area (position β and γ, Fig. 2A). A priory, it is not clear whether the Sn intercalation is mainly restricted to the evaporated region or by either surface diffusion or diffusion between the buffer layer and if the substrate the entire system is intercalated. To investigate a possible transition of the intercalated to the non-intercalated area, we have taken a set of AFM/KPFM data starting in the Sn intercalated center to a region well below the shadow mask (Figure 2A). Fig. 2 B/C shows AFM/KPFM data taken at the Sn deposited area of the sample at position α. The corresponding topography is, on the one hand, partly dominated by strong step bunching and, on the other hand, exhibiting a more gradual slope within the rest. The CPD also shows spatial variations. The LEED results, in combination with the sign and relative magnitude of the CPD differences, let us conclude that these areas correspond to mostly (1×1) reconstruction (lower CPD) and patches of $(\sqrt{3} \times \sqrt{3})$ (higher CPD). We find a strong spatial correlation of topography and CPD. A higher CPD value is found at the step bunched terraces, and a lower value is found at the gradual slope of the topography.

Well below the shadow mask at position γ, we expect to find the pristine buffer layer. AFM/KPFM data from this area can be seen in Fig. 2F/G. The topography shows a step-bunched surface. The CPD shows a nearly constant value on the different terraces with local signatures at the surface steps with varying values of up to 600 mV higher than the terraces.

We have taken a series of AFM/KPFM data going from the Sn covered area to the buffer layer region. While most data show either the signature of either the (1×1)/$(\sqrt{3} \times \sqrt{3})$ intercalated phase or the buffer layer, the dataset in Fig. 2 D/E shows a measurement with a clear spatial transition in the CPD. Two clearly distinct values separated by about 270 mV are found.

To judge this value, we have evaluated the difference in CPD of different material combinations. For details, see SI1. The CPD of (1×1) Sn intercalated graphene is about 400 mV higher than the CPD of graphene. The CPD difference between graphene and the unannealed buffer layer is about 640 mV. Moreover, we have calibrated our CPD values by direct comparison on gold. Based on all these values, the eventually estimated CPD difference between (1×1) Sn intercalated graphene and the buffer layer is about 240 mV, which is close to the above experimentally determined value of 270 mV. We therefore conclude that the left side of Fig. 2E with the high CPD is identified as (1×1) Sn intercalated graphene and the buffer layer on the right with a lower relative CPD.

In a next step, we have analyzed the spatial variation of CPD taken the topographic information at the transition region into account. The latter is dominated by surface steps ranging from 5 to 50 Å resulting from the miscut of the substrate and the step bunching during the growth process.

The spatial variation of the CPD signal aligns to a great extent with the step structure. On a length scale of several μm along a step, the drop of the CDP coincides with the steps. To illustrate this, Fig. S1 in the SI shows an overlay of the CPD with the topography. We find this kind of transitions down to bi-atomic steps (500 pm) due to the 4H SiC-substrate. From this observation, we conclude that surface steps represent a strong barrier for the intercalation process.

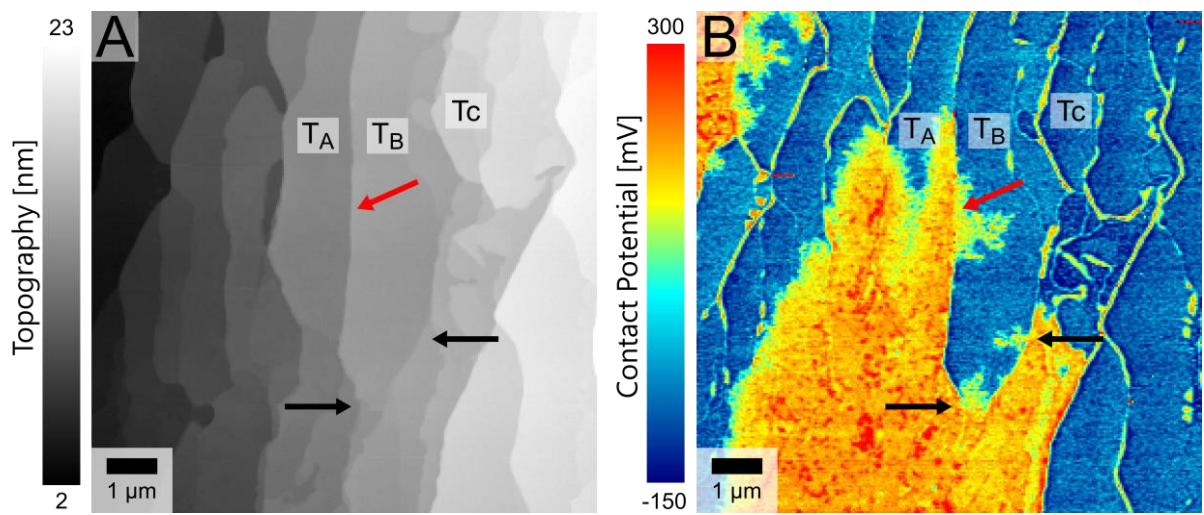

*Figure 3. The figure shows a zoom-in of figure 2D/E. (A) The AFM topography shows a step like structure. (B) shows the corresponding CPD-map. Three terraces are labeled with $T_A$-$T_C$. Arrows mark specific points of interest, where the Sn enters the next terrace with varying conditions dependent on the lateral position.*

To further investigate the tin transport across the steps we analyze the Sn distribution on the terrace next to a fully covered terrace. Our approach is based on the idea that the CPD can be used as a fingerprint to separate intercalated areas from the buffer layer. Figure 3 shows a region where the intercalation of such a terrace has started. The lower terrace $T_A$ is separated by a step of 2.5 nm from the higher one $T_B$, also the next step to the terrace $T_C$ is included. As the value of CPD is in-between the buffer layer's and fully intercalated areas, we conclude that the contrast is due to a partly intercalation by Sn.

The terrace $T_A$ is an example where the transition from a fully intercalated area to the pristine area can be found without steps involved. $T_B$ is a mainly pristine terrace. Intercalation starts at steps indicated by the red arrow in Fig. 3B. The feature fades out when proceeding on $T_B$. This is a strong indication that the transport across a step is not dominantly homogeneous along the step but is happening predominantly at specific sites within the step, referred to as pinholes. More complicated step geometries are also present and marked with black arrows at the lower part of the figure. The topography shows that in these areas of the surface, two different step edges meet and the tin is flowing to the terrace from these positions.

While the center area of $T_B$ has the CPD value of the buffer layer, a higher CPD value is found at the next step edge from $T_B$ to $T_C$. The decoration close to the step is exclusively located on the upper terrace and changes by up to 270 mV in CPD, relative to the pristine terrace. This value is strikingly close to the value of the fully Sn (1×1) intercalated terraces. Moreover, typically, a step is not decorated homogeneously, and not every step edge is decorated in that manner. Extended decorated areas show the same value as the fully (1×1) intercalated part on the left side of Fig. 3B. This indicates that the (1×1) intercalation starts at the steps.

To exclude the possibility that the decoration is due to an intrinsic step signature of the BL, we have investigated a pristine buffer layer sample. The overall lateral variation of the CPD looks alike, i.e., also showing a different CPD value at the step edges in contrast to the terraces. The severe difference to Fig. 3B is, that the CPD step signal of the reference sample is much higher, double than the value at the Sn intercalated region. For details, see SI3. From this, we conclude that it is, in fact, Sn that decorates the step edges.

The intercalation width at steps is up to 250 nm (that is well above the spatial resolution of our setup). The position of the decoration at the steps does not show a correlation with the position of pinholes, it shows a wide spread instead. This could be either attributed to a high mobility on the terrace or a high mobility of Sn at the steps. Combining the missing signature of intercalation on the terrace (e.g. $T_B$), we moreover conclude that steps are dominant nucleation sites.

Based on this assumption, we can investigate the intercalation process of a single terrace. Figure 4 shows an example where the right edge (step up) of a terrace is decorated, the left edge (step down) is still unaffected. Taking the idea that the absolute value of the CPD is a measure for the degree of Sn intercalation, the intercalation of this terrace proceeds from the top-right to the bottom-left part. Details on the evaluation can be found in SI4.

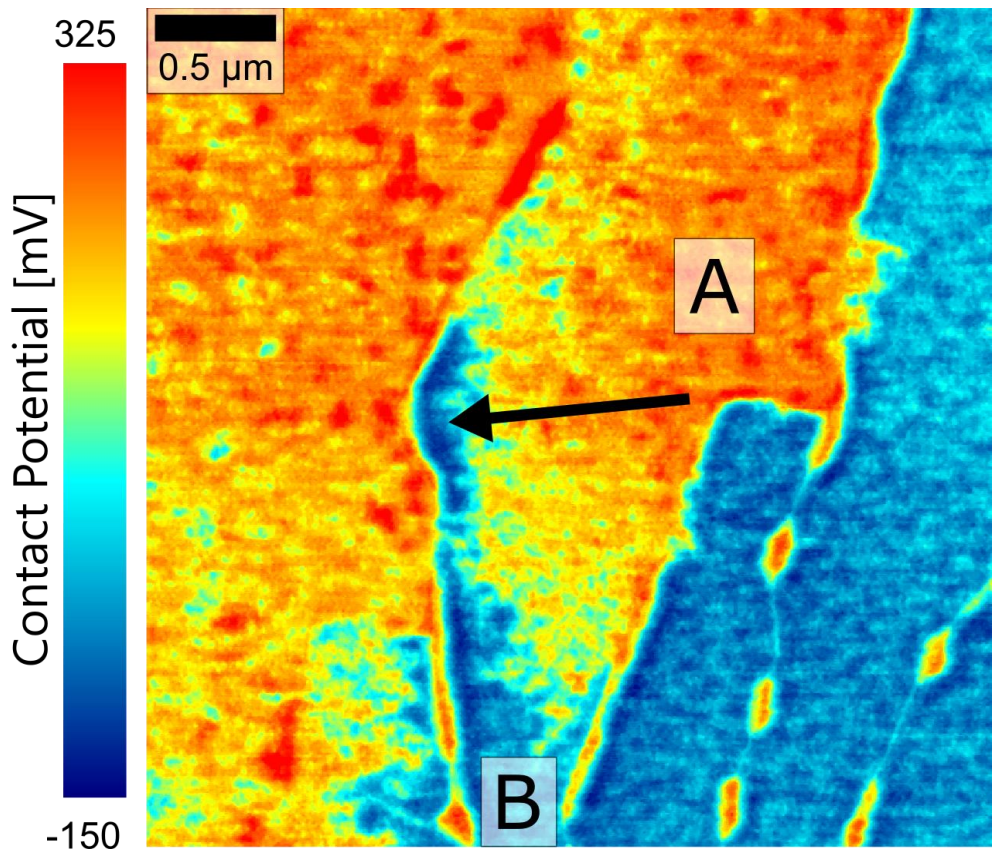

*Figure 4. CPD map showing a 2D intercalated area. Starting from an entire intercalated area (A), a gradual transition to a buffer layer (B) on a single terrace is observed. The CPD gradient suggests, that the grow on the terrace starts on the right side to the left.*

## Discussion

The main experimental results can be summarized as follows:

- Substrate steps are dominant atom transport barriers.
- Transport across a step is happening at local sites in the step.
- Transport on the terrace is fast and nucleation on the terrace is not found.
- The lower edge of steps are nucleation sites, at upper edges it was not observed.
- 2D interaction on the terrace starts from the nucleation sites at step up.

Based on this, we propose the following intercalation model of Sn and buffer layer graphene. In general, intercalation of Sn and buffer layer graphene starts as an exchange process of the buffer layer carbon atoms and the Sn atoms eventually forming the graphene sheet on top of an Sn-terminated SiC surface. While the buffer layer is characterized by a mixture of $sp^2$- and $sp^3$-bonds between the graphene atoms and the SiC substrate, graphene is $sp^2$ hybridized [15]. Nucleation sites of intercalation are areas where the Sn has mainly overtaken the role of C at the interface to SiC. The experimental results show that this happens at the bottom part of steps. Moreover, nucleation needs pinhole free steps, i.e. a significant diffusion barrier. Given the context of the step edges as potential barriers, the diffusion barrier can be interpreted with-in the concept of the Ehrlich-Schwoebel-Barrier [23]. It outlines the potential landscape with periodic potential on a terrace according to the surface reconstruction and a much higher binding potential barrier at steps, resulting in a lower probability overcoming the edges. Assuming a weakening of the barrier potential due to defects or a complex step geometry would lead to the observed pinholes (Fig. 5A).

This process, atom transport from the bottom to the top of a step, results in single or few Sn atoms on the next terrace. Hypothesizing, two options how the Sn transport from a pinhole to the next steps could happen. Firstly, the Sn diffuses via exchange processes with the carbon atoms of the buffer layer. Secondly, the Sn becomes an adatom on the buffer layer and diffuses as mobile species (Fig. 5 B,C, yellow particles). For the given sample, both cases result obviously in no nucleation sites on the terrace but "intercalation nucleation" at the bottom of the next step. This is similar to the transformation of the buffer layer to graphene also starting at the bottom of a step as the decomposition of the SiC substrate is enhanced. Within this model, the 2D intercalation starts at the nucleation site at the steps proceeding in the opposite direction (Fig. 5B).

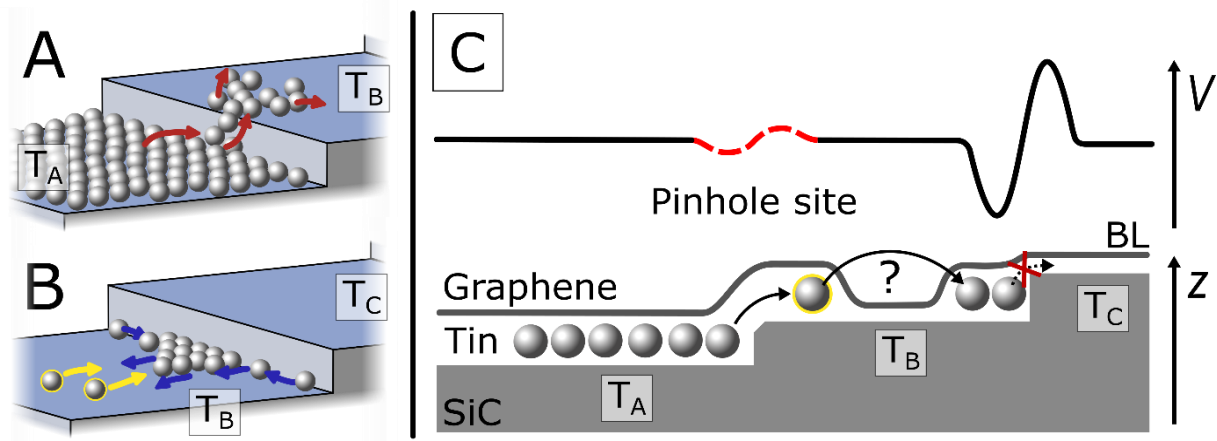

*Figure 5. The illustration of the model for the Sn intercalation processes described in the main text. A. Pinholes are relevant for the transport across steps acting as diffusion barriers. Note, the carbon layer that covers the tin and the surface of the SiC is implied. B. Intercalation nucleation starts at the bottom of the steps. C. The top part sketches the potential landscape, indicating the Schwoebel-Barrier and the breakdown at pinholes. The atom transport across the terrace from the pinhole to the next step edge could either be done as intercalated or on-surface atoms.*

We like to mention that the intercalation process strongly depends on the sample preparation conditions. Figure S4 in the supplement shows an example, where the overall sample preparation has been done at lower temperatures. The transition region in this case is dominated by an expected evolution to a diffusion front (for details see SI5).

## Conclusion

In this study, we conducted combined atomic force and kelvin probe force microscopy measurements in ambient conditions to study the intercalation process of Sn and buffer layer graphene. Spatially resolved contact potential difference maps are used to disentangle the different mechanisms of intercalation. Steps are identified as significant diffusion barriers; pin holes are found to transport Sn across a step. Intercalation nucleation starts at steps, resulting in a 2D growth against the macroscopic concentration gradient.

## Acknowledgment
We gratefully acknowledge the cooperation with AG Seyller from the Technische Universität Chemnitz. Christoph Lohse, who grew the used buffer layer on SiC samples and Peter Richter for discussing the manuscript.

## Financing Statement
Financial support of the Deutsche Forschungsgemeinschaft (DFG) through We1889/14-1, Te386/22-1 within FOR5242 and Te386/251 is acknowledged.

# References

[1] Chen S, Zhao D, Chen L, Liu G, Ding Y, Cao Y et al. Emerging Intercalation Cathode Materials for Multivalent Metal-Ion Batteries: Status and Challenges. Small Structures 2021;2(11).

[2] Stark MS, Kuntz KL, Martens SJ, Warren SC. Intercalation of Layered Materials from Bulk to 2D. Advanced materials (Deerfield Beach, Fla.) 2019;31(27):e1808213.

[3] Riedl C, Coletti C, Starke U. Structural and electronic properties of epitaxial graphene on SiC(0 0 0 1): a review of growth, characterization, transfer doping and hydrogen intercalation. J. Phys. D: Appl. Phys. 2010;43(37):374009.

[4] Mamiyev Z, Tegenkamp C. Exploring graphene-substrate interactions: plasmonic excitation in Sn-intercalated epitaxial graphene. 2D Mater. 2024;11(2):25013.

[5] Wundrack S, Momeni D, Dempwolf W, Schmidt N, Pierz K, Michaliszyn L et al. Liquid metal intercalation of epitaxial graphene: Large-area gallenene layer fabrication through gallium self-propagation at ambient conditions. Phys. Rev. Materials 2021;5(2).

[6] Rosenzweig P, Karakachian H, Link S, Küster K, Starke U. Tuning the doping level of graphene in the vicinity of the Van Hove singularity via ytterbium intercalation. Phys. Rev. B 2019;100(3).

[7] Gruschwitz M, Ghosal C, Shen T-H, Wolff S, Seyller T, Tegenkamp C. Surface Transport Properties of Pb-Intercalated Graphene. Materials (Basel, Switzerland) 2021;14(24).

[8] Schädlich P, Ghosal C, Stettner M, Matta B, Wolff S, Schölzel F et al. Domain Boundary Formation Within an Intercalated Pb Monolayer Featuring Charge-Neutral Epitaxial Graphene. Adv Materials Inter 2023;10(27).

[9] Herrera SA, Parra-Martínez G, Rosenzweig P, Matta B, Polley CM, Küster K et al. Topological Superconductivity in Heavily Doped Single-Layer Graphene. ACS nano 2024;18(51):34842–57.

[10] Wolff S, Tilgner N, Speck F, Schädlich P, Göhler F, Seyller T. Quasi-Freestanding Graphene via Sulfur Intercalation: Evidence for a Transition State. Adv Materials Inter 2024;11(2).

[11] Lin Y-R, Wolff S, Schädlich P, Hutter M, Soubatch S, Lee T-L et al. Vertical structure of Sb-intercalated quasifreestanding graphene on SiC(0001). Phys. Rev. B 2022;106(15).

[12] Link S, Forti S, Stöhr A, Küster K, Rösner M, Hirschmeier D et al. Introducing strong correlation effects into graphene by gadolinium intercalation. Phys. Rev. B 2019;100(12).

[13] Rosenzweig P, Starke U. Large-area synthesis of a semiconducting silver monolayer via intercalation of epitaxial graphene. Phys. Rev. B 2020;101(20).

[14] Jin L, Fu Q, Yang Y, Bao X. A comparative study of intercalation mechanism at graphene/Ru(0001) interface. Surface Science 2013;617:81–6.

[15]	Goler S, Coletti C, Piazza V, Pingue P, Colangelo F, Pellegrini V et al. Revealing the atomic structure of the buffer layer between SiC(0 0 0 1) and epitaxial graphene. Carbon 2013;51:249–54.

[16]	Wu S, Zhang Q, Yang H, Ma Y, Zhang T, Liu L et al. Advances in two-dimensional heterostructures by mono-element intercalation underneath epitaxial graphene. Progress in Surface Science 2021;96(3):100637.

[17]	Kruskopf M, Pakdehi DM, Pierz K, Wundrack S, Stosch R, Dziomba T et al. Comeback of epitaxial graphene for electronics: large-area growth of bilayer-free graphene on SiC. 2D Mater. 2016;3(4):41002.

[18]	Mamiyev Z, Tegenkamp C. Sn intercalation into the BL/SiC(0001) interface: A detailed SPA-LEED investigation. Surfaces and Interfaces 2022;34:102304.

[19]	Mamiyev Z, Balayeva NO, Ghosal C, Zahn DR, Tegenkamp C. Confinement induced strain effects in epitaxial graphene. Carbon 2025;234:120002.

[20]	Ghosal C, Ryee S, Mamiyev Z, Witt N, Wehling TO, Tegenkamp C. Electronic correlations in epitaxial graphene: Mott states proximitized to a relativistic electron gas; 2024.

[21]	Melitz W, Shen J, Kummel AC, Lee S. Kelvin probe force microscopy and its application. Surface Science Reports 2011;66(1):1–27.

[22]	Scheithauer U, Meyer G, Henzler M. A new LEED instrument for quantitative spot profile analysis. Surface Science 1986;178(1-3):441–51.

[23]	Schwoebel RL, Shipsey EJ. Step Motion on Crystal Surfaces. Journal of Applied Physics 1966;37(10):3682–6.

[24]	Mammadov S, Ristein J, Krone J, Raidel C, Wanke M, Wiesmann V et al. Work function of graphene multilayers on SiC(0001). 2D Mater. 2017;4(1):15043.

[25]	Kuzumoto Y, Kitamura M. Work function of gold surfaces modified using substituted benzenethiols: Reaction time dependence and thermal stability. Appl. Phys. Express 2014;7(3):35701.

[26]	Junge F, Auge M, Zarkua Z, Hofsäss H. Lateral Controlled Doping and Defect Engineering of Graphene by Ultra-Low-Energy Ion Implantation. Nanomaterials (Basel, Switzerland) 2023;13(4).

## Supplemental Information

SI1: Estimation of the contact potential transition

Due to spontaneous modifications of the tip, the absolute value of the contact potential difference of a given area changes. Therefore, areas of our inhomogeneous surface separated by more than the scan range of our AFM setup cannot be directly characterized with an absolute value.

To take this into account, we conducted a reference measurement on the surface of a single crystal of gold directly after measuring the transition region with the assumed pristine buffer layer. The results show that the work function of the buffer layer is very close to that of gold. The former published value of an unannealed BL sample in UHV reported a work function of around 4.8 eV [24]. The published work function values of gold vary, but are found to be in the same range [25]. Although not needed for the further discussion, this allows us to estimate the absolute work function values.

Evaluating the shift of the WF at the transition region, is possible in two steps.

Firstly, taking the WF of pristine graphene on a buffer layer into account, we have a value of 4.16 eV [24]. Due to the partial polarization through the substrate, the doping level of the graphene shifts to an n-type value of 400 meV above the Dirac-point. Knowing that the (1×1) reconstruction of tin reverses this effect back to neutral [4], one can now modify the WF of pristine graphene of 4.16 eV with 0.4 eV to get 4.56 eV for the combined system. That this shift of the doping level results in an equivalent shift of the WF was proven by [26] and also shown with the here used setup with a partly boron implanted graphene sample following the same procedure. Note, that the CPD is the negative of the relative work function of the given measurement.

Secondly, one can compare the difference of these values of the BL and the modified graphene layer to the difference of the WF in our data shown in Fig. 2 D/E, that is 0.27 eV. Comparing this to the difference of the published values of the buffer layer and the modified pristine graphene with (1×1) tin reconstruction one gets 4.8 eV – 4.56 eV ≈ 0.24 eV, that fits very well to our result of the difference of the transition of WF in sign and magnitude.

With this, we interpret the given transition in Fig. 2D/E as the one from a tin intercalated graphene sheet to the pristine BL.

SI2: Contour Plot

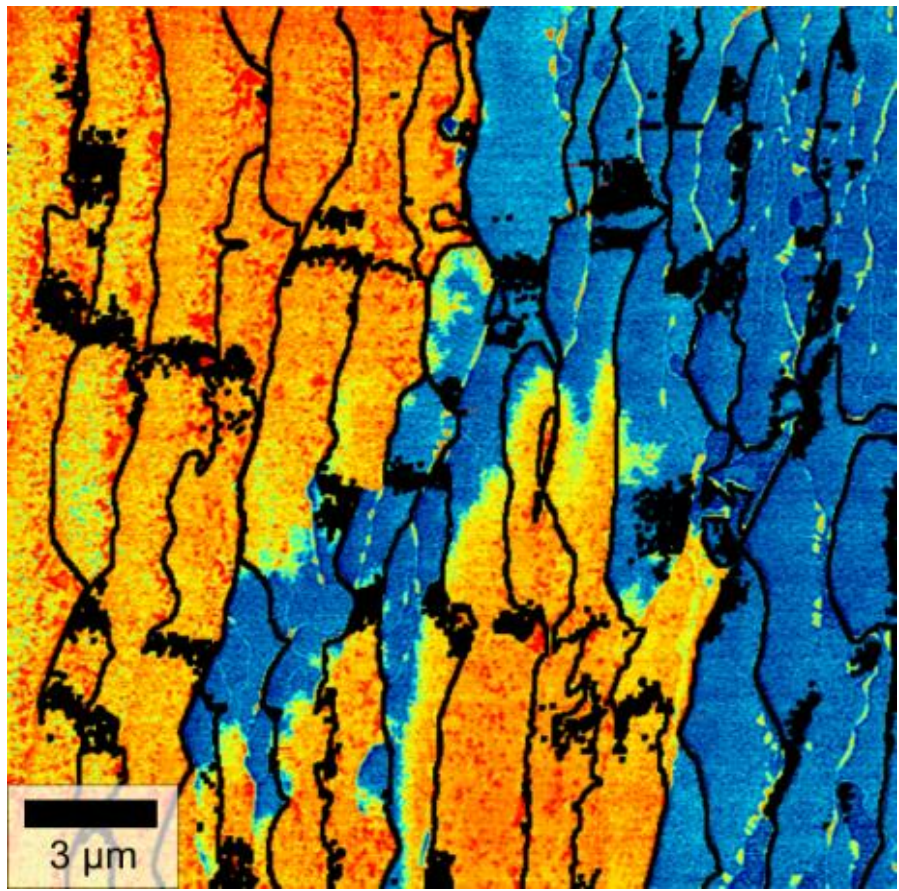

*Figure S1. This shows an overlay of the contours of the dominant surface steps with the CPD regarding the data set in Fig. 2D/E.*

Figure S1 shows an overlay of the contours of substrate steps (>2 nm) shown in Fig. 2D with the corresponding CPD map Fig.1E. The border of the high CPD areas are clearly correlated to the step contours over the majority of the transition region. Therefore, this confirms, that the steps are acting as major diffusion barriers.

SI3: Comparing Surface Signatures Intercalated and Pristine

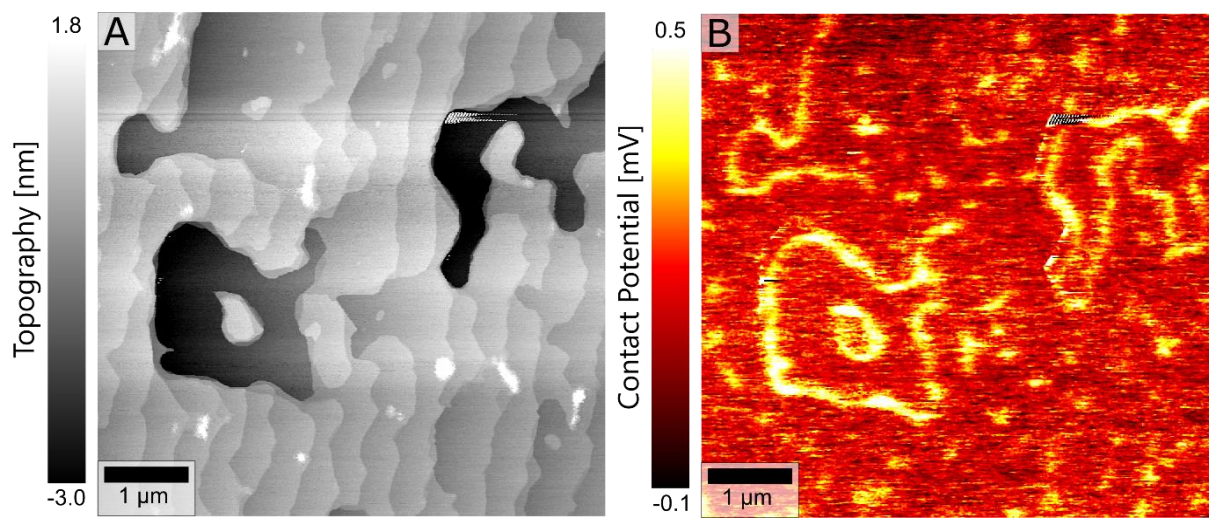

*Figure S2. AFM topography and corresponding CPD map of a pristine buffer layer sample. High CPD signatures are correlated to the position of the steps in the topography.*

The dataset shown in Fig. S2 was taken on a pristine BL sample, that was grown with same parameter as the sample discussed in the rest of the paper. A corresponds to the topography, B to the simultaneously measured CPD-map at that position. It shows, similar to figure 3F/G, a pronounced CPD signature at steps. As the difference with up to 600 mV is much larger than the one within Sn intercalated areas (Fig. 2C), we conclude that the step signatures in figure 2E and 3 are caused by Sn intercalation. This is supported by the fact that the CPD contrast of the intercalated sample at steps increases deeper below the shadow mask (Fig. 2G), where presumable less Sn should be found.

## SI4: Comparing Surface Signatures Intercalated and Pristine

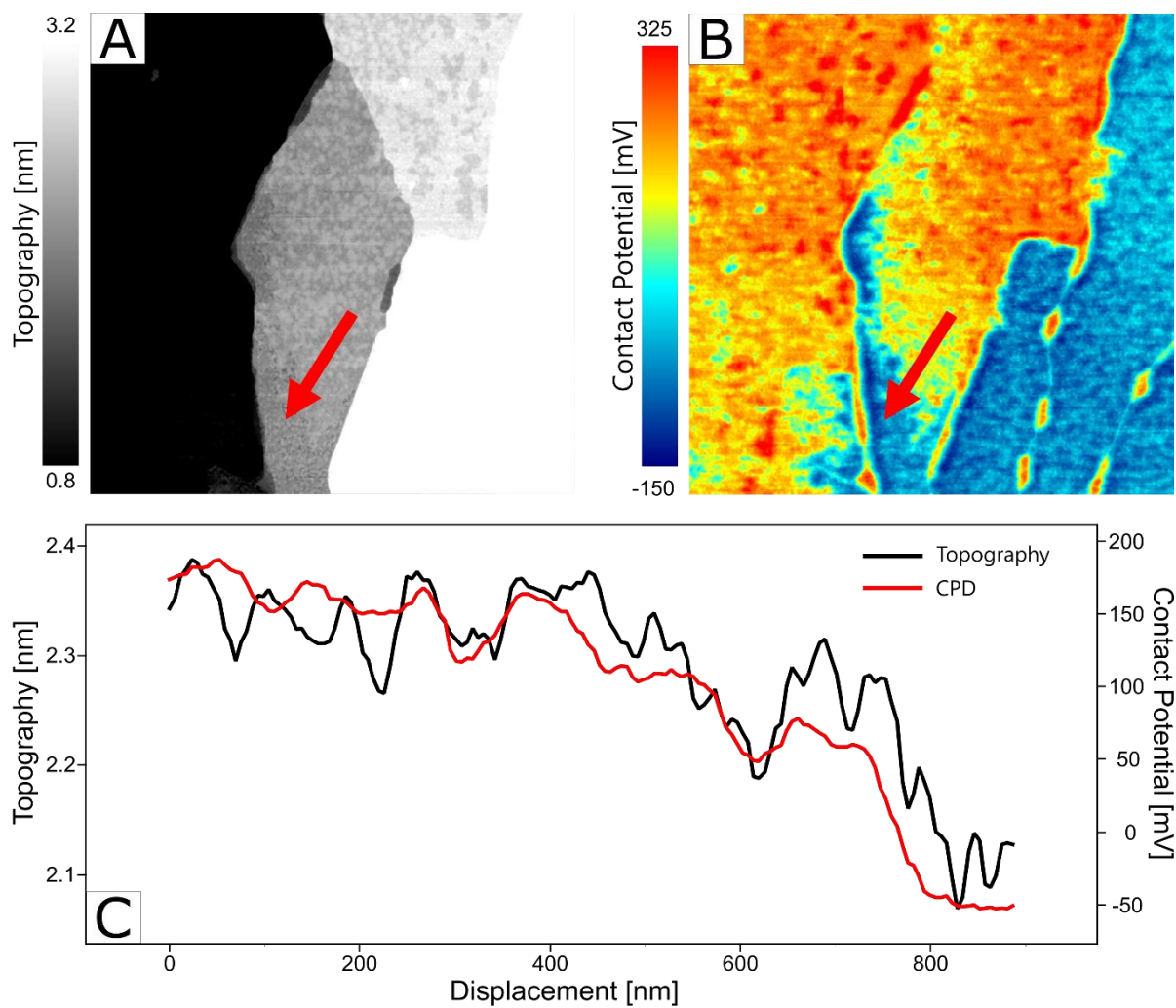

*Figure S3. Analysis of the 2D growth on a single terrace's gradual transition from an intercalated region to the pristine BL for a corresponding data set in A and B, and the section along the red arrow in A and B, respectively, in C.*

Figure S3 shows a transition from intercalated to non-intercalated pristine BL on one single terrace. The topography is shown in S3A, the according CPD-map in Fig. S3B and the section along the arrow is plotted in Fig. S3C. There is a strong correlation of the topography to the CPD-value. This is not a discrete transition, as one could expect, given a single Sn monolayer is forming in the intercalated region. Please note, that this is different to the sample that was prepared with lower temperatures in SI 5, where a much more discrete transition is found.

SI5: Sn Intercalation at lower temperature sample preparation

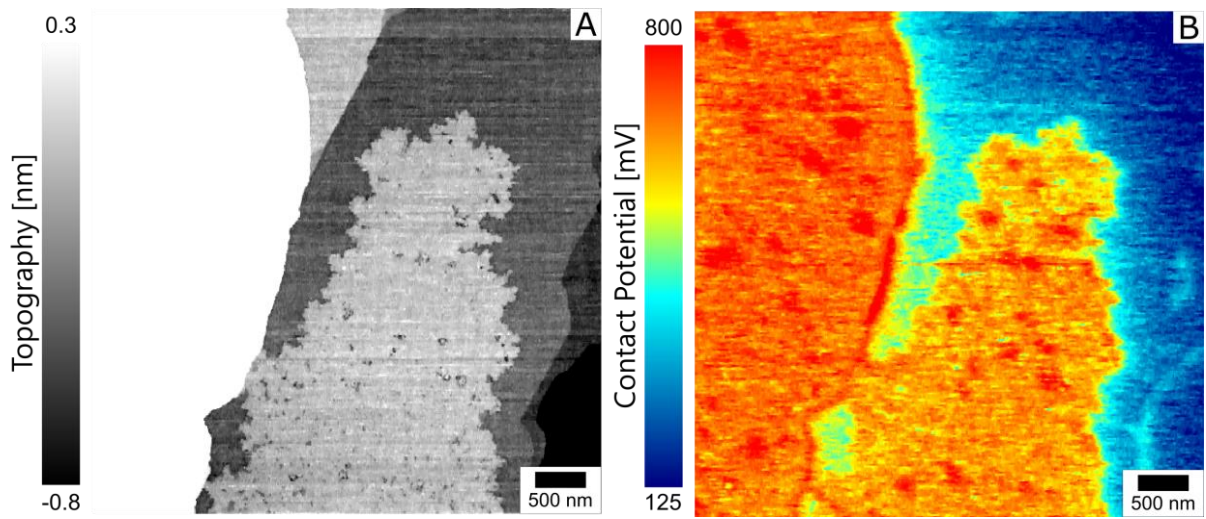

*Figure S4. The figure shows the intercalation to pristine BL transition with a different sample, that was prepared with a lower temperature. The diffusion front on the terrace was formed without a decoration of any step edges and seems to be mostly directed at the terrace itself.*

Preparation at a lower temperature ($\leq$ 800°C) results in an entirely different transition region from the intercalated area to the pristine buffer layer. This can be observed in S4, that shows the topography (A) and CPD-map (B) of a likewise prepared sample, with the difference that the temperature of the intercalation procedure was lower.

There are two things noticeable: Firstly, the transition is less gradual. Secondly, the propagation seems to favor an opposite behavior, not being dominated by the step edges, but to be mostly happening on the terrace itself.